\documentclass[10pt,a4paper,twoside]{article}
\usepackage{indentfirst}
\usepackage{graphicx}
\usepackage{amsgen,amsfonts,amssymb,amsbsy}

\newenvironment{resum}{\begin{quote}\small}{\end{quote}}

\begin{document}

\thispagestyle{plain} 

\begin{center}
{\LARGE \textsf{\textbf{Spacetime Foam and Vacuum Energy}}}

\bigskip

\textbf{Remo Garattini}

\textsl{Facolt\`{a} di Ingegneria, Universit\`{a} di Bergamo}

\textsl{Viale Marconi, 5, 24044 Dalmine (Bergamo) Italy}

\textsl{E-mail: Garattini@mi.infn.it }
\end{center}

\medskip

\begin{resum}
A simple model of spacetime foam, made by spherically symmetric wormholes,
with or without a cosmological term is proposed. The black hole area
quantization and its consequences are examined in this context. We open the
possibility of probing Lorentz symmetry in this picture.
\end{resum}

\bigskip

\section{Introduction}

The term Spacetime Foam was introduced for the first time by J.A.
Wheeler at the end of the fifties to indicate that quantum
fluctuations come into play at the Planckian scale, changing
topology and metric \cite{J.A. Wheeler}. Apart the pioneering path
integral approach to Spacetime Foam, considered by S.W. Hawking
\cite{S.W. Hawking}, only recently the subject has been widely
reconsidered in the context of brane physics \cite{EMN}, spin foam
models \cite {BCHT} and on a phenomenological ground
\cite{phenomenological}.
We will briefly introduce a different approach based on a large N$_{w}$%
-wormhole semi-classical approximation. This model is motivated by
the results on the computation of the Casimir energy computed on
some wormhole backgrounds of spherical symmetry (Schwarzschild,
Schwarzschild-de Sitter, Schwarzschild-Anti-de Sitter and
Reissner-Nordstr\"{o}m metrics) which have as a reference space
the presumed ground state (Minkowski, de Sitter and Anti-de Sitter
metrics) \cite{RIJMPD,RCQG1,RCQG2,RIJMPA2}. The formulation of the
Casimir effect in general is synthesized by the Eq.
\begin{equation}
E_{Casimir}\left[ \partial \mathcal{M}\right] =E_{0}\left[ \partial \mathcal{%
M}\right] -E_{0}\left[ 0\right] ,
\end{equation}
where $E_{0}$ is the Zero Point Energy (Z.P.E.) and $\partial \mathcal{M}$
is a boundary. It is immediate to see that the Casimir energy involves a
vacuum subtraction procedure and since this one is related to Z.P.E., we can
extract information on the ground state. In concrete terms, we compute the
following quantity
\[
E\left( wormhole\right) =E\left( no-wormhole\right)
\]
\begin{equation}
+\Delta E_{no-wormhole}^{wormhole}{}_{|classical}+\Delta
E_{no-wormhole}^{wormhole}{}_{|1-loop},  \label{i0}
\end{equation}
representing the total energy computed to one-loop in a wormhole
background. $E\left( no-wormhole\right) $ is the reference space.
$\Delta E_{no-wormhole}^{wormhole}{}_{|classical}$ is the
classical energy difference between the wormhole and no-wormhole
configuration stored in the boundaries and finally $\Delta
E_{no-wormhole}^{wormhole}{}_{|1-loop}$ is the quantum correction
to the classical term. Usually, the last term in the examined
metrics exhibits an unstable mode. In the Coleman language, we are
perturbing the false vacuum \cite{Coleman}. Instead of rejecting
such a configuration, we take into examination a large wormholes
number.

\section{Large N$_{w}$-wormhole approach to Spacetime foam}

We consider $N_{w}$\ non-interacting wormholes in a semiclassical
approximation and assume that there exists a covering of $\Sigma $\ such
that $\Sigma =\cup _{i=1}^{N_{w}}\Sigma _{i}$, with $\Sigma _{i}\cap \Sigma
_{j}=\emptyset $\ when $i\neq j$. Each $\Sigma _{i}$\ has the topology $%
S^{2}\times R^{1}$\ with boundaries $\partial \Sigma _{i}^{\pm }$.
Furthermore, we assume the existence of a bifurcation surface $S_{0}$, which
is the case for the wormhole background considered. On each surface $\Sigma
_{i}$ the energy stored in the boundaries is zero because we assume that on
each copy of the single wormhole there is symmetry with respect to each
bifurcation surface\footnote{%
This can be proved in terms of quasilocal energy.}. In this context, the
total energy contribution comes only by quantum fluctuations. Moreover, the
instability appearing in the single wormhole case is eliminated in the
multi-wormhole picture. As a consequence, the regularized total energy -
Casimir energy - assumes the form, at its minimum,
\begin{equation}
\Delta E_{N_{w}}\left( \bar{x}\right) =-N_{w}\frac{V}{64\pi ^{2}}\frac{%
\Lambda ^{4}}{e}  \label{a4}
\end{equation}
valid even for the Schwarzschild-Anti-de Sitter case\footnote{%
The case of the Schwarzschild-de Sitter case gives for the Z.P.E. the
following expression
\[
\Delta E_{N_{w}}\left( \bar{x}\right) =-N_{w}\frac{V}{32\pi ^{2}}\frac{%
\Lambda ^{4}}{e}.
\]
}.

\subsection{Some consequences of a foamy spacetime}

Here we will consider some applications of the foam model. One consequence
is a natural quantization process in terms of the wormholes constituents. In
particular, the area of a black hole is examined in this context. The area
is measured by the quantity
\begin{equation}
A\left( S\right) =\int_{S}d^{2}x\sqrt{\sigma }.
\end{equation}
$\sigma $ is the two-dimensional determinant coming from the induced metric $%
\sigma _{ab}$ on the boundary $S$. The evaluation of the mean value of the
area
\begin{equation}
A\left( S\right) =\frac{\left\langle \Psi _{F}\left| \hat{A}\right| \Psi
_{F}\right\rangle }{\left\langle \Psi _{F}|\Psi _{F}\right\rangle }=\frac{%
\left\langle \Psi _{F}\left| \widehat{\int_{S}d^{2}x\sqrt{\sigma }}\right|
\Psi _{F}\right\rangle }{\left\langle \Psi _{F}|\Psi _{F}\right\rangle },
\end{equation}
is computed on the following state
\begin{equation}
\left| \Psi _{F}\right\rangle =\Psi _{1}^{\perp }\otimes \Psi _{2}^{\perp
}\otimes \ldots \ldots \Psi _{N_{w}}^{\perp }.
\end{equation}
Suppose to consider the mean value of the area $A$ computed on a given
\textit{macroscopic} fixed radius $R$. On the basis of our foam model, we
obtain $A=\bigcup\limits_{i=1}^{N}A_{i}$, with $A_{i}\cap A_{j}=\emptyset $
when $i\neq j$. Thus $A=4\pi l_{p}^{2}N\alpha $. $\alpha $ represents how
each single wormhole is distributed with respect to the black hole in terms
of the area. When compared with the Bekenstein area spectrum proposal, $%
\alpha $ is fixed to $\ln 2/\pi $ and \cite{Area}
\begin{equation}
M=\frac{\sqrt{N}}{2l_{p}}\sqrt{\frac{\ln 2}{\pi }},
\end{equation}
namely the Schwarzschild black hole mass is \textit{quantized} in terms of $%
l_{p}$. The level spacing of the transition frequencies is
\begin{equation}
\omega _{0}=\Delta M=\left( 8\pi Ml_{p}^{2}\right) ^{-1}\ln 2.
\end{equation}
An interesting aspect comes when we consider Schwarzschild-Anti-de Sitter
wormholes as foam constituents. Indeed, the level spacing of the transition
frequencies is
\begin{equation}
\omega _{0}=\Delta M_{S}^{AdS}=\left( 8M_{S}l_{p}^{2}\right) ^{-1}\frac{9\ln
2}{16\pi }=\Delta M\frac{9}{16}
\end{equation}
that it means that for a given Schwarzschild black hole of mass $M$, the
S-AdS foam representation gives smaller frequencies. In this respect, we
have found a possibility to understand which constituents form our foam
model. The inclusion of a charge is straightforward and leads to
\begin{equation}
Q^{2}=\sqrt{N_{2}}\left( \sqrt{N_{1}}-\sqrt{N_{2}}\right) \alpha l_{p}^{2},
\end{equation}
where $N_{2}$ is the wormholes number used for the covering of the RN black
hole area. We immediately see that from the above equation we have $%
N_{1}\geq N_{2}$, where the equality corresponds to the vanishing
charge. Moreover we choose $N_{1}$ and $N_{2}$ in such a way that
$Q^{2}=\alpha l_{p}^{2}q$, $q=0,1,2,\ldots $. This means that
\begin{equation}
\sqrt{N_{2}}\left( \sqrt{N_{1}}-\sqrt{N_{2}}\right) =q.
\end{equation}
When $q=0$, we recover the Schwarzschild case, namely $N_{2}=N_{1}$. On the
other hand, when we consider $N_{1}=4q$, we get $N_{2}=q$ corresponding to
the extreme case. This leads to the quantization of the Reissner-Nordstr\"{o}%
m black hole mass
\begin{equation}
M=\frac{\sqrt{\alpha }}{2l_{p}}\sqrt{N_{2}}\left( 1+\frac{q}{N_{2}}\right) .
\end{equation}
When we consider a transition in mass with a fixed charge, on the
level spacing we get
\begin{equation}
\omega _{0}=\Delta M=\frac{\partial M}{\partial r_{+}}\frac{dr_{+}}{dN}%
\Delta N=\frac{\pi }{A}\left( r_{+}-r_{-}\right) \alpha   \label{p49}
\end{equation}
with $\Delta N=1$, which vanishes in the extreme limit. Finally, for the de
Sitter geometry, we write
\begin{equation}
S=N\ln 2=\frac{3\pi }{l_{p}^{2}\Lambda _{c}}=\frac{A}{4l_{p}^{2}}=\frac{%
N4\pi l_{p}^{2}}{4l_{p}^{2}}=N\pi ,
\end{equation}
that is
\begin{equation}
\frac{3\pi }{l_{p}^{2}N\ln 2}=\Lambda _{c}.  \label{Lambda}
\end{equation}
An interesting aspect appears when we put numbers in Eq.$\left( \ref{Lambda}%
\right) $. When $N=1$, the foam system is highly unstable and the
cosmological constant assumes the value, in order of magnitude, of $\Lambda
_{c}\sim 10^{38}GeV^{2}$. However the system becomes stable when the whole
universe has been filled with wormholes of Planckian size and this leads to
the huge number $N=10^{122}$ corresponding to the value of $\Lambda _{c}\sim
10^{-84}GeV^{2}$ which is the order of magnitude of the cosmological
constant of the space in which we now live.

\subsection{Lorentz invariance? }

Recently, a lot of interest has been devoted to the problem of
Lorentz invariance at high energies. There are several reasons to
suspect that Lorentz invariance may be only a low energy symmetry.
This possibility is suggested by the ultraviolet divergences of
local quantum field theory, as well as by tentative results in
various approaches to quantum gravity and string theory
\cite{Lorentz invariance}. One signal of breaking of such a
symmetry should be given by a modification of the dispersion
relation deviating from
\begin{equation}
E^{2}(p)=p{^{2}}+{m^{2},}
\end{equation}
where for simplicity we have considered the case of a massive scalar field.
One possible simple distortion can be characterized at low energies by an
expansion with integral powers of momentum,
\begin{equation}
E^{2}=m^{2}+Ap^{2}+Bp^{3}/E_{0}+Cp^{4}/E_{0}^{2}+O(p^{5}).
\end{equation}
$E_{0}$ is the ``quantum gravity'' scale (Planck energy $\simeq $ $10^{19}$
GeV). The dispersion relation is not Lorentz invariant. It can only hold in
one reference frame. One possible ``\textit{preferred reference frame''}
could be represented by space-time foam\footnote{%
See also Ref.\cite{CraneSmolin}, where Lorentz symmetry has been explicitly
broken.}. In our foam model we have the possibility to verify the above
dispersion relation, which can be also used to probe the validity of this
picture.

\setlength{\itemsep}{-0.6 ex}

\end{document}